\documentclass[twocolumn,showpacs,preprintnumbers,floatfix]{revtex4}

\usepackage{makeidx}
\usepackage{amssymb}
\usepackage{graphicx}

\begin{document}

\title{\textbf{Quasiperiodic graphs: structural design, scaling and
entropic properties}}
\author{B. Luque$^1$, F. J. Ballesteros$^2$, A. M. N\'u\~{n}ez$^1$ and A.
Robledo$^3$}
\affiliation{$^1$ Dept. Matem\'{a}tica Aplicada y Estad\'{i}stica. ETSI Aeron\'{a}uticos, Universidad Polit\'{e}cnica de Madrid, Spain.\\
$^2$  Observatori Astron\`{o}mic. Universitat de Val\`{e}ncia, Spain.\\
$^3$ Instituto de F\'{\i}sica y Centro de Ciencias de la Complejidad, Universidad Nacional Aut\'{o}noma de M\'{e}xico, Mexico.}

\begin{abstract}
A novel class of graphs, here named quasiperiodic, are constructed via application of the Horizontal
Visibility algorithm to the time series generated along the
quasiperiodic route to chaos. We show how the hierarchy of mode-locked
regions represented by the Farey tree is inherited by their associated
graphs. We are able to establish, via Renormalization Group (RG) theory, the
architecture of the quasiperiodic graphs produced by irrational winding numbers
with pure periodic continued fraction. And finally, we demonstrate that the RG
fixed-point degree distributions are recovered via optimization of a suitably
defined graph entropy.
\end{abstract}

\maketitle

Quasiperiodicity is observed along time evolution in nonlinear dynamical
systems \cite{chaos, steve, hilborn} and also in the spatial arrangements of
crystals with forbidden symmetries \cite{shechtman, fractals}. These two
manifestations of quasiperiodicity are rooted in self-similarity and are
seen to be related through analogies between incommensurate quantities in time
and spatial domains \cite{fractals}. Here we point out that quasiperiodicity
can be visualized in a third way: in the graphs generated when
the Horizontal Visibility (HV) algorithm \cite{PNAS, PRE} is applied to the
stationary trajectories of the universality class of low-dimensional
nonlinear iterated maps with a cubic inflexion point, as represented by the
circle map \cite{fractals}.

The idea of mapping time series into graphs has been presented
in recent works \cite{zhang06, Kyriakopoulos07, xu08, donner10, donner11, donner11-2,
campanharo11} where different approaches have been
developed. In particular, the period-doubling bifurcation
cascade has been analyzed in the light of the HV formalism \cite{plos, Chaos}
and a complete set of graphs, called Feigenbaum graphs, that encode the
dynamics of all stationary trajectories of unimodal maps has been provided.
The Feigenbaum scenario is one of the three well-known routes to reach chaos in
low-dimensional dissipative systems (along with the intermittency route and
the quasiperiodicity route) \cite{chaos, steve, hilborn}. In this Letter we
characterize the structural, scaling and entropic properties of the graphs
obtained when the HV formalism is applied to the quasiperiodic routes to
chaos. As we shall see, a Renormalization Group (RG) treatment of such
graphs is the instrument that grants access to our main results.

\begin{figure}[tbp]
\includegraphics[width=1.0\columnwidth,angle=0,scale=1.0]{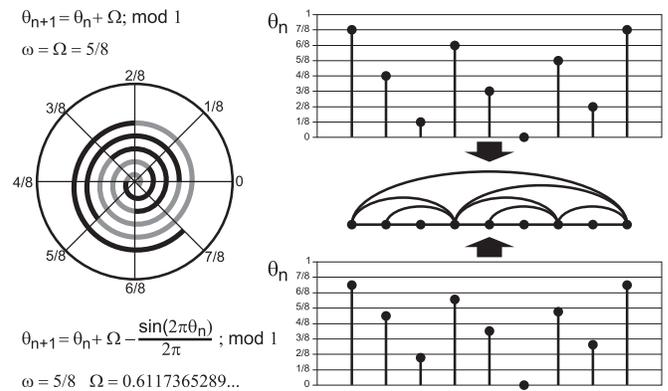}
\caption{{\protect\small  Examples of two standard circle map periodic series with dressed winding number
$\omega=5/8$, $K=0$ (top) and $K=1$ (bottom). As can be observed, the order of visits on the circle and the relative values of $\theta_n$ remain invariant and the associated
HV graph is therefore the same in both cases.} }
\label{mapping}
\end{figure}

We briefly recall that the standard
circle map \cite{chaos, steve, hilborn} is the one-dimensional iterated map
given by:
\begin{equation}
\theta _{t+1}=f_{\Omega ,K}(\theta _{t})=\theta _{t}+\Omega -\frac{K}{2\pi }
\sin (2\pi \theta _{t}),\;\text{mod}\;1,  \label{circle1}
\end{equation}
representative of the general class of nonlinear circle maps:
$\theta _{t+1}=f_{\Omega ,K}(\theta _{t})=\theta _{t}+\Omega +K\cdot g(\theta _{t}),\;\text{mod}\;1$,
where $g(\theta)$ is a periodic function that fulfills $g(\theta+1)=g(\theta)$.
The HV graphs obtained for this family of maps exhibit universal properties that without loss of generality we explain in the next paragraphs in terms of the standard circle map.

The dynamical variable $0\leq \theta _{t}<1$ can be interpreted as a measure of the angle
that specifies the trajectory on the unit circle, the control parameter $
\Omega $ is the so-called \emph{bare winding number}, and $K$ is a
measure of the strength of the nonlinearity.
The \emph{dressed winding number} for the map is defined as the limit
of the ratio: $\omega \equiv \lim_{t\rightarrow \infty }(\theta _{t}-\theta
_{0})/t$ and represents an averaged increment of $\theta_t$
per iteration. For $0\leq K \leq1$ trajectories are periodic (locked motion) when
the corresponding dressed winding number $\omega (\Omega )$ is a rational
number $p/q$ and quasiperiodic when it is irrational. The winding
numbers $\omega (\Omega )$ form a devil's staircase which
makes a step at each rational number $\omega =p/q$ and remains constant for
a range of $\Omega $. For $K=1$ ($\emph{critical}$ circle map) locked motion
covers the entire interval of $\Omega $ leaving only a multifractal subset
of $\Omega $ unlocked.

\begin{figure*}[tbp]
\includegraphics[width=1.0\columnwidth,angle=0,scale=1.8]{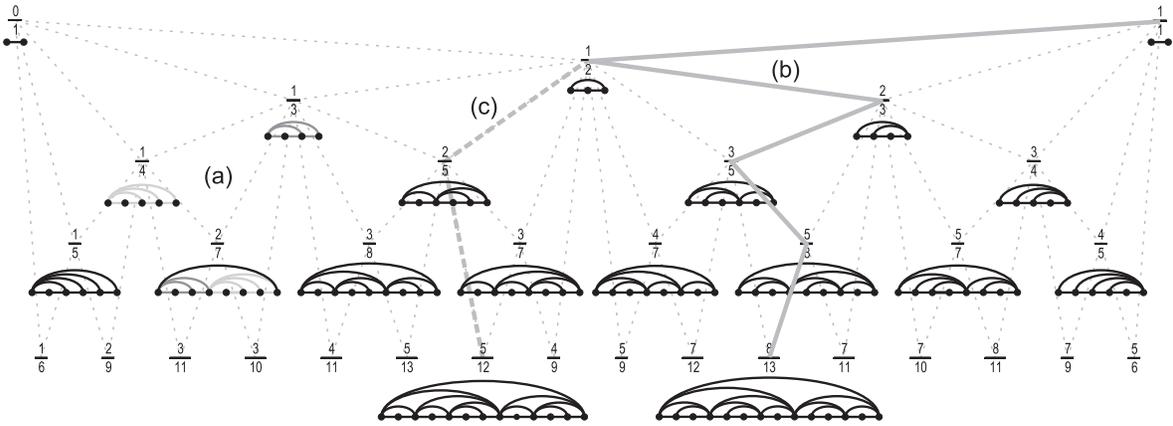}
\caption{{\protect\small Six levels of the Farey tree and the periodic
motifs of the graphs associated with the corresponding rational fractions $p/q$
taken as dressed winding numbers $\protect\omega$ in the circle map (for space reasons only
two of these are shown at the sixth level). (a) In order to show how graph concatenation works, we have
highlighted an example using different grey tones on the left side: as $1/3>1/4$, $G(1/3)$
is placed on the left, $G(1/4)$ on the right and their extremes are connected to an additional
link closing the motif $G(2/7)$.
(b) Five steps in the Golden ratio route, $b=1$ (thick solid line);
(c) Three steps in the Silver
ratio route, $b=2$ (thick dashed line).} }
\label{farey}
\end{figure*}

The resulting hierarchy of mode-locking steps at $K=1$ can be conveniently represented by a
Farey tree which orders all the irreducible rational numbers $p/q\in[0,1]$ according
to their increasing denominators $q$. In the devil's staircase, $\omega(\Omega)$,
the width of the steps (intervals where $\omega$ is
constant) becomes smaller when the denominator $q$ increases. Furthermore,
if we have two steps with winding numbers $p/q$ and
$p^{\prime }/q^{\prime }$, the largest step between them has a winding
number $(p+p^{\prime })/(q+q^{\prime })$, which is also the irreducible rational number
with the smallest denominator. Thus, the Farey tree also orders all mode-locking steps with
$\omega =p/q$ in the circle map according to their decreasing widths \cite{hao}.

The HV algorithm assigns each datum $\theta_i$ of a
time series $\{\theta_{i}\}_{i=1,2,...}$ to a node $i$ in its associated HV graph, and
$i$ and $j$ are two connected nodes if $\theta_{i}, \theta_{j}>\theta_{n}$ for all $n$ such that $i<n<j$.
Without loss of generality, we apply the HV algorithm to the superstable orbits of the critical circle map ($K=1$)
with an irreducible rational number $\omega(\Omega)= p/q$. Thus, the associated time series always contains
$\theta_0=0$ as one of its values and has period $q$ (cf. \cite{chaos,steve,hilborn}).
The associated HV graph is a periodic repetition of a motif with $q$ nodes, $p$ of which have connectivity $k=2$. (Observe that
$p$ in the map indicates the number of turns in the circle to complete a period).
For $K \leq 1$, the order of visits of positions
in the attractors and their relative values remain invariant for a locked region with $\omega=p/q$ \cite{hao},
such that the HV graphs associated with them are the same. In fig. \ref{mapping} we present an example where
the first and the last node in the motif correspond to the largest value in the attractor.

\begin{figure}[tbp]
\includegraphics[width=1.0\columnwidth,angle=0,scale=1.0]{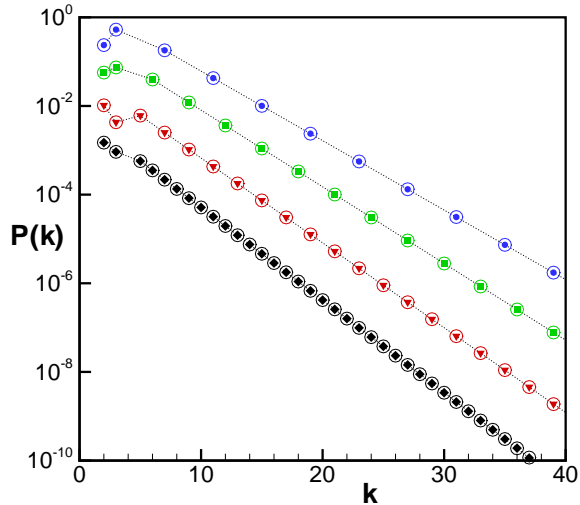}
\caption{{\protect\small Empty circles stand for theoretical degree
distributions of quasiperiodic graphs. Filled values have been obtained
by direct application of the HV algorithm to critical circle map series of $10^{6}$ values.
Distributions have been shifted from each other to enhance visualization. From down-up:
(a) $\omega_{\infty}=[\bar{1}]$, $\Omega=0.606661...$.
(b) $\omega_{\infty}=[\bar{2}]$, $\Omega=0.418864...$;
$\omega_{\infty}=[\bar{3}]$, $\Omega=0.323873...$;
$\omega_{\infty}=[\bar{4}]$, $\Omega=0.271502...$.} }
\label{figure2}
\end{figure}

In fig. \ref{farey} we depict the associated HV periodic motifs for each $p/q$ in the Farey tree.
We observe straightforwardly that the graphs can be constructed
by means of the following inflation process: let $p/q$ be a Farey
fraction with `parents' $p^{\prime }/q^{\prime }<p^{\prime \prime
}/q^{\prime \prime }$, i.e., $p/q=(p^{\prime }+p^{\prime \prime })/(q^{\prime
}+q^{\prime \prime })$. The `offspring' graph $G(p/q)$ associated with $\omega =p/q$,
can be constructed by the concatenation $G(p^{\prime \prime }/q^{\prime
\prime }) \oplus G(p^{\prime }/q^{\prime })$ of the graphs of its parents.
By means of this recursive construction we can systematically explore the structure of every graph
along a sequence of periodic attractors leading to quasiperiodicity.
A standard procedure to study the quasiperiodic route to chaos is fixing $K=1$
and selecting an irrational number $\omega _{\infty }\in[0,1]$. Then,
a sequence $\omega _{n}$ of rational numbers approaching $\omega _{\infty }$ is taken.
This sequence can be obtained through successive
truncations of the continued fraction expansion of $\omega _{\infty }$. The
corresponding bare winding numbers $\Omega (\omega _{n})$ provide attractors
whose periods grow towards the onset of chaos, where the period of the
attractor must be infinite. A well-studied case is the sequence of rational
approximations of $\omega _{\infty }=\phi^{-1} =(\sqrt{5}-1)/2\simeq 0.6180...$,
the reciprocal of the Golden ratio, that yields winding numbers $\{\omega
_{n}=F_{n-1}/F_{n}\}_{n=1,2,3...}$ where $F_{n}$ is the Fibonacci number
generated by the recurrence $F_{n}=F_{n-1}+F_{n-2}$ with $F_{0}=1$ and
$F_{1}=1$. The first few steps of this route are shown in fig. \ref{farey}(b):
$\omega_1=1/1,\ \omega_2=1/2,\ \omega_3=2/3,\ \omega_4=3/5,\ \omega_5=5/8...,\ \omega_6=8/13...$ .
Within the range $\Omega(F_{n-1}/F_{n})$ one observes trajectories of
period $F_{n}$ and, therefore, this route to
chaos consists of an infinite family of periodic orbits with increasing
periods of values $F_{n}$, $n\rightarrow \infty$.
If we denote by $G_{\phi^{-1}}(n)$ the graph associated to
$\omega _{n}=F_{n-1}/F_{n}$ in the Golden ratio route, it is easy to prove
that the associated connectivity distribution $P(k)$ for
$G_{\phi^{-1}}(n)$ with $n\geq 3$ and $k\leq n+1$
is $P_{n}(2)=F_{n-2}/F_{n}$, $P_{n}(3)=F_{n-3}/F_{n}$, $P_{n}(4)=0$ and
$P_{n}(k)=F_{n-k+1}/F_{n}$. In the limit $n\rightarrow \infty $ the
connectivity distribution at the accumulation point $G_{\phi^{-1}}(\infty)$,
the quasiperiodic graph at the onset of chaos, takes the form
\begin{equation}
P_{\infty }(k)=\left\{
\begin{array}{ll}
1-\phi^{-1}  & k=2 \\
2\phi^{-1} -1 & k=3 \\
0 & k=4 \\
\phi^{1-k} & k\geq 5.%
\end{array}%
\right.   \label{golden_distrib}
\end{equation}%
Figure \ref{figure2}(a) shows that the theoretical degree distribution of the
quasiperiodic graph for the route described above is in perfect agreement with the same quantity
obtained by applying the HV algorithm to a circle map time series with
a dressed winding number $\omega_{\infty}=\phi^{-1}$. The procedure explained
for the Golden ratio can be repeated for the `time reversed sequence':
$\{\omega_{n}=F_{n-2}/F_{n}=1-F_{n-1}/F_{n}\}_{n=1,2,3,..}$. In this case the ratio
converges to $1-\phi^{-1} $ in the limit $n\rightarrow \infty $. The connectivity
distributions of the graphs $\{G_{1-\phi^{-1} }(n)\}_{n=1,2,3,..}$ are the same
as in the Golden ratio route because these graphs are symmetric mirror versions
of the former (we use the term `time'
because the `time reverse' of a graph
from a series generated by a clockwise rotation in the circle map
corresponds to the graph from the same but counterclockwise
rotation).

The previous results can be interpreted
through a suitably-defined Renormalization Group (RG) transformation. We
proceed as in previous works \cite{plos, Chaos} and define the RG graph operation $%
\mathcal{R}$ as the coarse-graining of every couple of adjacent nodes where
one of them has degree $k=2$ into a block node that inherits the links of
the previous two nodes. If we continue with the case of the Golden ratio, we
first note that $\mathcal{R}\{G_{\phi^{-1} }(n)\}=G_{1-\phi^{-1} }(n-1)$ and $\mathcal{%
R}\{G_{1-\phi^{-1} }(n)\}=G_{\phi^{-1} }(n-1)$, so the RG flow alternates between the
two mirror routes. If we define the operator `time reverse' by
$\overline{G}_{\phi^{-1} }(n)\equiv G_{1-\phi^{-1} }(n)$, the
transformation becomes $\overline{\mathcal{R}}\{G_{\phi^{-1}
}(n)\}=G_{\phi^{-1} }(n-1)$ and $\overline{\mathcal{R}}\{G_{1-\phi^{-1}
}(n)\}=G_{1-\phi^{-1} }(n-1)$. Repeated application of $\overline{\mathcal{R}}$
yields two RG flows that converge, for $n$
finite, to the trivial fixed point $G_{0}$ (a graph with $P(2)=1$).
The accumulation points $n\rightarrow \infty$, the quasiperiodic graphs, act as nontrivial
fixed points of the RG flow: $\overline{\mathcal{R}}\{G_{\phi^{-1} }(\infty
)\}=G_{\phi^{-1} }(\infty )$ and $\overline{\mathcal{R}}\{G_{1-\phi^{-1} }(\infty
)\}=G_{1-\phi^{-1} }(\infty )$.

The above RG procedure works only in the case of the Golden ratio route. (As a counterexample look at the so-called Silver ratio route shown in fig. \ref{farey}c).
To extend the above formalism to other irrational numbers,
we develop the following explicit algebraic version of $\mathcal{R}$ and apply it to the Farey
fractions associated with the graphs,
\begin{equation}
R\bigg(\frac{p}{q}\bigg)=\left\{
\begin{array}{ll}
R_{1}\big(\frac{p}{q}\big)=\frac{p}{q-p} & \mbox{if }\frac{p}{q}<\frac{1}{2}
\\
R_{2}\big(\frac{p}{q}\big)=1-\frac{q-p}{p} & \mbox{if }\frac{p}{q}>\frac{1}{2%
},%
\end{array}%
\right.
\end{equation}%
along with the algebraic analog of the `time reverse' operator
$\overline{R}(x)=1-R(x)$.
Observe that along the Golden ratio route fractions are always greater than
$1/2$ and we can therefore renormalize this route by making
\begin{equation}
\overline{R}\bigg(\frac{F_{n-1}}{F_{n}}\bigg)=\overline{R_{2}}\bigg(\frac{%
F_{n-1}}{F_{n}}\bigg)=\frac{F_{n-2}}{F_{n-1}},
\end{equation}%
whose fixed-point equation $\overline{R}(x)=x$ is
$ x^{2}+x-1=0$,
with $\phi^{-1}$ a solution of it.

A straightforward generalization of this scheme is obtained by considering the routes
$\{\omega _{n}=F_{n-1}/F_{n}\}_{n=1,2,3...}$ with $F_{n}=bF_{n-1}+F_{n-2}$,
$F_{0}=1$, $F_{1}=1$ and $b$ a natural number. It is easy to see that
$\lim_{n\rightarrow \infty }F_{n-1}/F_{n}=(-b+\sqrt{b^{2}+4})/2$, which is a
solution of the equation $x^{2}+bx-1=0$. Interestingly,
all the positive solutions of the above family of quadratic
equations happen to be positive quadratic irrationals in $[0,1]$ with pure periodic continued fraction representation:
$\phi^{-1} _{b}=[b,b,b,...]=[\bar{b}]$ ($b=1$ corresponds to the Golden route). Every $b>1$ fulfills the condition
$F_{n-1}/F_{n}<1/2$, and, as a result, we have
\begin{equation}
R\bigg(\frac{F_{n-1}}{F_{n}}\bigg)=R_{1}\bigg(\frac{F_{n-1}}{F_{n}}\bigg)=%
\frac{F_{n-1}}{(b-1)F_{n}+F_{n-2}}.
\end{equation}%
The transformation $R_{1}$ can only be applied $(b-1)$ times before the
result turns greater than $1/2$, so the subsequent application of $R$ followed
by reversion yields
\begin{equation}
\overline{R^{(b)}}\bigg(\frac{F_{n-1}}{F_{n}}\bigg)=\overline{R_{2}}\bigg[%
R_{1}^{(b-1)}\bigg(\frac{F_{n-1}}{F_{n}}\bigg)\bigg]=\frac{F_{n-2}}{F_{n-1}}.
\end{equation}%
It is easy to demonstrate by induction that
\begin{equation}
R_{1}^{(b-1)}(x)=\frac{x}{1-(b-1)x},
\end{equation}%
whose fixed-point equation $\overline{R^{(b)}}(x)=\overline{R_{2}}[R_{1}^{(b-1)}(x)]=x$
leads in turn to $x^{2}+bx-1=0$, with $\phi^{-1}_{b}$ a solution of it.
We can proceed in an analogous way for the symmetric case $\omega
_{n}=1-(F_{n-1}/F_{n})$ but, as the sense of the inequalities for $1/2$ is
reversed, the role of the operators $R_{1}$ and $R_{2}$ must be exchanged.

The previous result indicates that graphs must be renormalized via
$\overline{\mathcal{R}^{b}}\{G_{\phi^{-1} _{b}}(n)\}=G_{\phi^{-1} _{b}}(n-1)$.
Again, the iteration of this process yields two RG flows that converge to
the trivial fixed point $G_{0}$ for $n$ finite. The quasiperiodic graphs, reached as accumulation points ($%
n\rightarrow \infty $), act as nontrivial fixed points of the RG flow since $%
\overline{\mathcal{R}^{b}}\{G_{\phi^{-1} _{b}}(\infty )\}=G_{\phi^{-1} _{b}}(\infty)$.

We observe that for fixed $b\geq2$, and
from the construction process illustrated in fig. \ref{farey}(a),
it can be deduced that $P_{\infty}(2)=\phi^{-1} _{b}$,
$P_{\infty}(3)=1-2\phi^{-1} _{b}$ and $P_{\infty}(k\neq bn+3)=0,\forall n \in \mathbb{N}$. $P_{\infty}(k=bn+3),\ n \in \mathbb{N}$ can be obtained from the condition of RG fixed-point invarance of the
distribution, as it implies a balance equation $P_{\infty}(k)=\phi^{-1}_b P_{\infty}(k+b)$
whose solution has the form of an exponential tail. The degree distribution
$P_{\infty}(k)$ for this quasiperiodic graphs is therefore
\begin{equation}
P_{\infty}(k)=\left\{
\begin{array}{ll}
\phi^{-1} _{b} & k=2 \\
1-2\phi^{-1} _{b} & k=3 \\
(1-\phi^{-1} _{b})\phi _{b}^{(3-k)/b} & k=bn+3,\;n\in \mathbb{N} \\
0 & \mathrm{otherwise.}%
\end{array}%
\right.
\label{metallic_distrib}
\end{equation}
A perfect agreement between theoretical and numerical results for some examples can be
observed in fig. \ref{figure2}(b).

Notably, all the RG flow directions and fixed points described
above can be derived directly from the information contained in the degree
distribution via optimization of the graph entropy functional
$H=-\sum_{k=2}^{\infty }{P(k)\log{P(k)}}$. The optimization is for a fixed $b$ and takes into account
the constrains: $P(2)=\phi^{-1} _{b}$, $P(3)=1-2\phi^{-1} _{b}$, maximum possible mean
connectivity $\langle k\rangle=4$ \cite{Chaos} and
$P(k)=0\ \forall k\neq bn+3,\ n\in \mathbb{N}$. The degree distributions $P(k)$
that maximize $H$ can be proven to be exactly the connectivity distributions of equations
\ref{golden_distrib} and \ref{metallic_distrib}
for the quasiperiodic graphs at the accumulation points found above. 
This establishes a functional relation between the fixed points of the RG flow and the extrema of $H$ as it was verified for
the period-doubling route \cite{plos,Chaos}.

We have demonstrated the capability of the HV algorithm
for transforming into graph language the universal properties of the route
to chaos via quasiperiodicity in low-dimensional nonlinear dynamical systems.
The outcome is a novel type of graph
architecture where the motifs are the building blocks with which quasiperiodicity
is expressed recursively via concatenation.
Significantly, the HV formalism leads to analytical expressions for the degree
distribution, a function that in all mode-locking regions is essentially
exponential. The networks' scaling properties can be formulated in terms of
an \emph{ad hoc} RG transformation for which the nontrivial graph fixed points
capture the features of the quasiperiodic accumulation points. As we have
seen, it is through the properties of the RG transformation presented above
that the relevant details of the quasiperiodic graphs studied are determined.
This class represents all the quasiperiodic attractors reached when irrational
winding numbers with pure periodic continued fractions are used as
dressed winding numbers. Furthermore, a graph entropy is introduced via the
degree distribution and its optimization reproduces the RG fixed points.

By means of the HV algorithm, we have found a
connection between pure periodic continued fractions and
the degree distribution of their associated quasiperiodic graphs. It seems
feasible to generalize our results beyond to periodic
continued fractions or any irrational with a pattern in its
continued fraction. Finally, it has not escaped our notice that, as we have a one-to-one correspondence
between graphs and rational numbers, a possible graph algebra can be explored.

BL and AN acknowledges support from FIS2009-13690 and S2009ESP-1691 (Spain); FB from  
AYA2006-14056, CSD2007-00060, and AYA2010-22111-C03-02 (Spain); AR from CONACyT \& DGAPA (PAPIIT
IN100311)-UNAM (Mexico).

\end{document}